\documentclass[intlimits,twoside,a4paper]{article}

\usepackage[cp1251]{inputenc}
\usepackage[eqsecnum]{cmpj3}

\issue{2022}{25}{4}{43706}
\doinumber{10.5488/CMP.25.43706}

\title[Phase transitions in ferroelectrics]%
{Phase transitions in ferroelectric domain walls%
}
\author[E. A. Eliseev, M. D. Glinchuk, L. P. Yurchenko, P. Maksymovych, A.~N.~Morozovska]{E. A. Eliseev\orcid{0000-0001-8124-8857}\refaddr{label1},
M. D. Glinchuk\orcid{0000-0002-5317-3082}\refaddr{label1},
L. P. Yurchenko\orcid{0000-0002-0345-9971}\refaddr{label1},
P. Maksymovych\orcid{0000-0003-0822-8459}\refaddr{label2},
A.~N.~Morozovska\orcid{0000-0002-8505-458X}\refaddr{label3}\thanks{Corresponding author: \email{anna.n.morozovska@gmail.com}.}}
\addresses{
\addr{label1} Institute for Problems of Materials Science, National Academy of Sciences of Ukraine, Krjijanovsky 3, 03142 Kyiv, Ukraine
\addr{label2} Center for Nanophase Materials Sciences, Oak Ridge National Laboratory, Oak Ridge, TN 37831
\addr{label3} Institute of Physics, National Academy of Sciences of Ukraine, 46, pr. Nauky, 03028 Kyiv, Ukraine
}


\date{Received June 21, 2022, in final form September 12, 2022}

\begin{document}

\maketitle

\begin{abstract}
Despite multiple efforts, there exist many unsolved fundamental problems related with detection and analysis of internal polarization structure and related phase transitions in ferroelectric domain walls. Their solution can be very important for the progress in domain wall nanoelectronics and related applications in advanced memories and other information technologies. Here, we theoretically study the features of phase transitions in the domain walls, which are potentially detectable by the scanning probe capacitance microwave microscopy. The finite element modelling based on the Landau-Ginzburg-Devonshire theory is performed for the capacitance changes related with the domain wall motion in a multiaxial ferroelectric BaTiO$_3$.
%
%
\printkeywords
%
\end{abstract}

\section{Introduction}


Since its appearance and until now, nanoscale ferroics (ferromagnets, ferroelectrics, ferroelastics) have been the main object of fundamental research on the physical nature of long-range order of polar, magnetic, and structural properties~\cite{Gli13,Hli19,Gru19}. The leading role is played by the emergence of a domain structure of long-range order parameters, such as electric polarization, magnetization, and (anti)ferrodistortion, and its interaction with the surface of nanoferroic~\cite{Wan19,Gli19}. In nanoferroics, the role of the surface increases significantly with a decrease in their size and very often begins to dominate in comparison with the bulk contribution~\cite{Kal18}. Moreover, this trend is not only due to the distinctive behavior of all properties near the surface but also due to the presence of the developed gradients of order parameters from the surface to the bulk of the nanoferroic~\cite{Gli13}. The development of the gradients largely generates and is generated by the appearance of internal electric and elastic fields near the surface of a ferroic.

Recently, it was predicted theoretically that a decrease in the correlation-gradient polarization energy leads to a significant increase in the polarization gradient, which in turn leads to spontaneous bending of uncharged domain walls in thin multiferroic films~\cite{Eli19} and in ferroelectric nanoparticles~\cite{Eli18}. Such walls can form meandering~\cite{Eli19} and/or labyrinthine~\cite{Eli18} structures. Later, similar structures were discovered experimentally in thin BiFeO$_3$ and Pb(Zr$_{0.4}$Ti$_{0.6}$)O$_3$ films by the high-resolution scanning transmission electron microscopy (HR-STEM)~\cite{Han19} and piezoresponse force microscopy (PFM)~\cite{Nah20} methods, respectively, and corroborated by ab initio calculations~\cite{Nah20}. HR-STEM, PFM and conductive atomic force microscopy (C-AFM) experiments registered an enhanced conductivity of nominally uncharged domain walls in ferroelectrics and multiferroics~\cite{Mak12,Bal12,Vas12}. Non-Ising and chiral ferroelectric domain walls were predicted theoretically~\cite{Gu14} and were revealed by nonlinear optical microscopy~\cite{Che17}. Despite multiple efforts, there exist fundamental problems related with the observation of the internal polarization structure and the related phase transitions in the domain walls. Their solution can be very important for the progress in domain wall nanoelectronics~\cite{Cat12,Sha22} and in the related advanced applications.

To contribute to the problems solution, here we theoretically study the features of phase transitions in ferroelectric domain walls, which are potentially observable using the scanning probe capacitance microwave microscopy~\cite{Tse16,Wu17,Chu20}. The finite element modelling (FEM) and analytical calculations based on the Landau-Ginzburg-Devonshire (LGD) theory are performed for the capacitance changes related with the domain wall motion in a multiaxial ferroelectric multiaxial ferroelectric BaTiO$_3$.

\section{Problem statement}

\begin{figure}[htb]
\centerline{\includegraphics[width=0.85\textwidth]{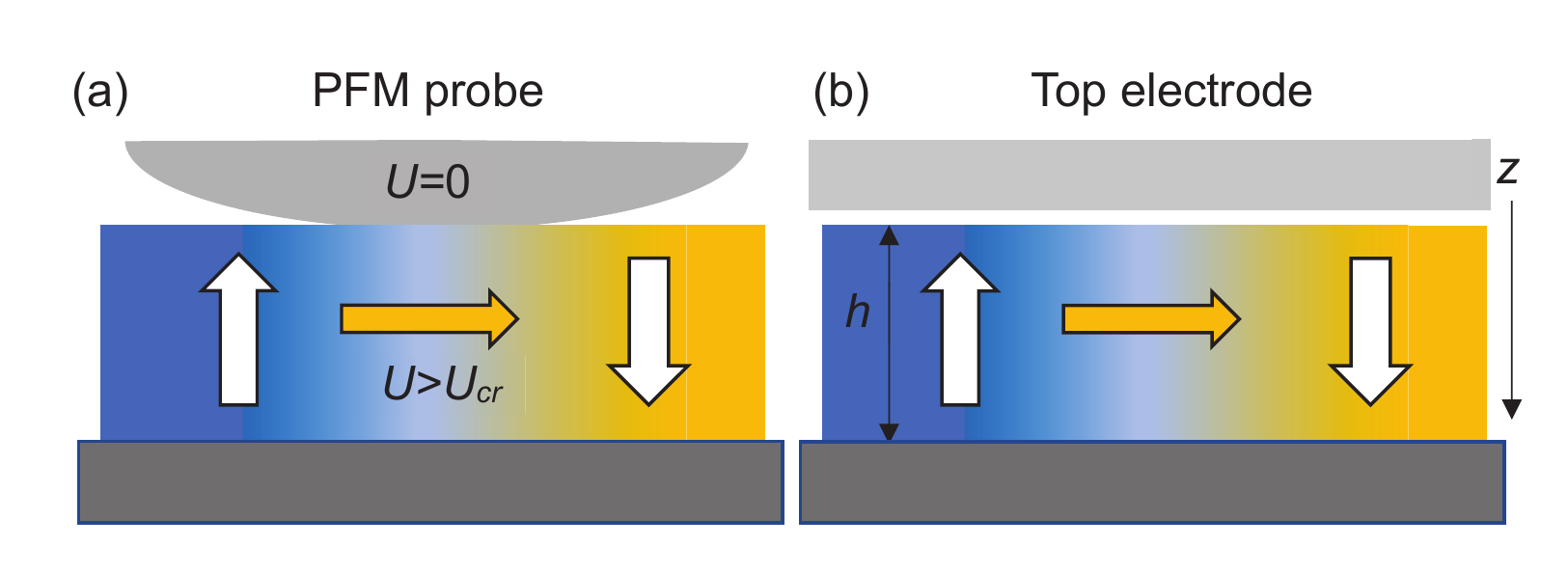}}
\caption{(Colour online) Ferroelectric thin film with a domain structure in the probe (a) and planar capacitor geometry (b).} \label{fig-smp1}
\end{figure}

Let us consider the features of the capacitance of a thin-film ferroelectric capacitor in a scanning probe geometry, with or without a domain structure. The control parameter is the temperature of the ferroelectric, which changes leads to the changes of the domain structure period, the width and orientation of the walls. Specifically, the magnitude and angles between the various components of polarization at the walls significantly depends on temperature and undergoes phase transitions. Typical parameters of the probe are the contact radius $\sim$50--100~nm (in fact, this is a free parameter), and the thickness of the ferroelectric film $\sim$10--20~nm [see figure~\ref{fig-smp1} (a)].

If the probe is modelled by a flat electrode (for example by a disk), then an ultrathin (less than 0.1~nm thick) dielectric gap between the PFM probe and the ferroelectric surface is required to stabilize the domain structure [see figure~\ref{fig-smp1} (b)]. The domain structure is unstable without the gap, and the system transits into a single-domain state at rather low voltages. However, even the presence of an ultrathin gap leads to the fact that a significant part of the applied voltage drops across it. To minimize this effect, the gap can be filled with a liquid dielectric having a high dielectric permittivity or can be replaced with a dead layer.

The electric polarization of the film is calculated from three coupled nonlinear Euler-Lagrange (EL) equations for each of the three components of the vector $P_i$, taking into account the model lattice potential (in fact, lattice pinning). These equations are derived from the variation of LGD thermodynamic potential, with respect to $P_i$. These EL equations must be accomplished with electrostatic and elastic problems. The pinning value determines the critical electric field required for the walls to move, and different models of pinning exist. As a rule, the critical field can be estimated analytically using several models. Suzuki-Ishibashi (S-I) model~\cite{Ish79} can be used for the threshold field determination. The activation filed, that determines the nucleation kinetics, can be defined within Miller-Weinreich~\cite{Mil60} or Burtsev-Chervonobrodov (B-C) models~\cite{Bur82} modified by Rappe et al.~\cite{Shi07} and Aravind et al.~\cite{Ara10}, respectively, for consideration of the phenomena related with polarization gradient and depolarization effects at the wall.

Note that the critical field of the wall motion depends on many factors~\cite{Cho08,Per13}, such as the structure (single-component and multicomponent) of the wall~\cite{Mor16}, which in turn is determined by the temperature and thickness of the film, by the bending of the wall and its mechanical clamping determined by the gradient of the applied electric field, electrostriction and flexoelectric couplings~\cite{Tag16,Wang19}.

For numerical modelling, we implement the simplest situation, when the lattice pinning can be modelled by a 3D quasi-harmonic potential that leads to the internal field ${E_i} \cong {E_0}\sin \bigl( {\vec{k} \vec x  + {\psi _0}} \bigr)$ in the right-hand side of the EL equations. Here, $\vec k$ is the reciprocal lattice vector, or its integral multiple.

The linear capacitance of the system can be calculated in two ways: as the second derivative of its electrostatic energy with respect to the applied electric voltage, or as the derivative of the total charge at the electrodes of the capacitor with respect to the voltage applied between the electrodes. In the case of an exact solution and linear capacity, as can be seen, both methods give the same answer. The first method turns out to be easier for numerical modelling, while the second method is sometimes more useful for analytical calculations.

\section{Capacitance of a single domain film}

For the model situation shown in figure~\ref{fig-smp1}~(b), except in the case of a single-domain polarization distribution $P_3(z)$, electric potential distribution $\varphi \left( z \right)$, charge density on the electrodes $\sigma$ and total system capacitance $C$ are equal to
\begin{eqnarray}
	\label{phi-pe}
	{\varphi _{FE}}\left( z \right) &=& \frac{U}{{1 + ({\varepsilon _b}/{\varepsilon _g})(d/h)}}\left( {1 - \frac{z}{h}} \right) + \frac{1}{{{\varepsilon _0}{\varepsilon _b}}}\Bigg[ {{\overline P }_3}\frac{{h - z}}{{1 + ({\varepsilon _b}/{\varepsilon _g})(d/h)}} \Bigg. \nonumber \\ 
	&-& \Bigg. \mathop \int_z^h {P_3}\left( {{z'}} \right)\rd{z'} \Bigg], \quad 0 < z < h,
\end{eqnarray}
\begin{align}
\label{phi-g}
{\varphi _g}\left( z \right) = \frac{U}{{1 + ({\varepsilon _b}/{\varepsilon _g})(d/h)}}\left( {1 - \frac{{{\varepsilon _b}}}{{{\varepsilon _g}}}\frac{z}{h}} \right) - \frac{{{{\overline P }_3}}}{{{\varepsilon _0}{\varepsilon _g}}}\frac{{z + d}}{{1 + ({\varepsilon _b}/{\varepsilon _g})(d/h)}},\quad  - d < z < 0,
\end{align}
\begin{align}
\label{sigma}
\sigma \left( { - d} \right) &= \frac{U}{{1 + ({\varepsilon _b}d/{\varepsilon _g}h)}}\frac{{{\varepsilon _0}{\varepsilon _b}}}{h} + \frac{{{{\overline P }_3}}}{{1 + ({\varepsilon _b}d/{\varepsilon _g}h)}}, \nonumber \\ 
 \sigma \left( h \right) &=  - \frac{U}{{1 + ({\varepsilon _b}d/{\varepsilon _g}h)}}\frac{{{\varepsilon _0}{\varepsilon _b}}}{h} - \frac{{{{\overline P }_3}}}{{1 + ({\varepsilon _b}d/{\varepsilon _g}h)}},
\end{align}
\begin{align}
\label{Ci}
C = S\frac{{\rd \sigma }}{{\rd U}} = S\left( \frac{{{\varepsilon _0}{\varepsilon _b}{\varepsilon _g}}}{{{\varepsilon _g}h + {\varepsilon _b}d}} + \frac{{{\varepsilon _g}h}}{{{\varepsilon _g}h + 
{\varepsilon _b}d}}\frac{{\rd{{\overline P }_3}}}{{\rd U}} \right),
\end{align}
where $h$ is the thickness of the ferroelectric film located at $0 < z < h$, $d$ is the thickness of the ultra-thin gap located at $- d < z < 0$. $S$ is the surface area of the electrodes. The upper surface of the film corresponds to $z = 0$, and the probe electrode under the applied voltage $U$ is located in the plane $z = - d$. The bottom electrode $z = h$ is grounded. The relative background permittivity of the ferroelectric film is $\varepsilon_b$, and the relative dielectric constant of the gap is $\varepsilon _g$. The average value of the spontaneous polarization normal component is defined as 
${\overline P_3} = ({1}/{h}) \int_0^h {P_3}\left( {{z'}} \right)\rd{z'}$. 
In the approximate equality~(\ref{Ci}), $C \approx S( {\varepsilon _0}{\varepsilon _b}{\varepsilon _g})/({\varepsilon _g}h + {\varepsilon _b}d)$, and the value ${\overline P _3}$ is regarded voltage-independent, and in this case, the capacitance is indifferent to structural phase transitions. Thus, only the behavior of the second term, which is ${\varepsilon _g}hS/({\varepsilon _g}h + {\varepsilon _b}d)\left( \rd{{\overline P }_3} /\rd U \right)$, can be an indicator of the transition from the paraelectric phase to the tetragonal, then to the orthorhombic and rhombohedral phases, which occurs under the temperature decrease. These transitions will correspond to the jumps in the relative dielectric susceptibility of a single-domain ferroelectric film ${\chi _f}\left( T \right)$, because $\rd{{\overline P }_3} / \rd U \sim {\chi _f}\left( T \right)$ in the linear approximation.

\begin{figure}[htb]
\centerline{\includegraphics[width=0.60\textwidth]{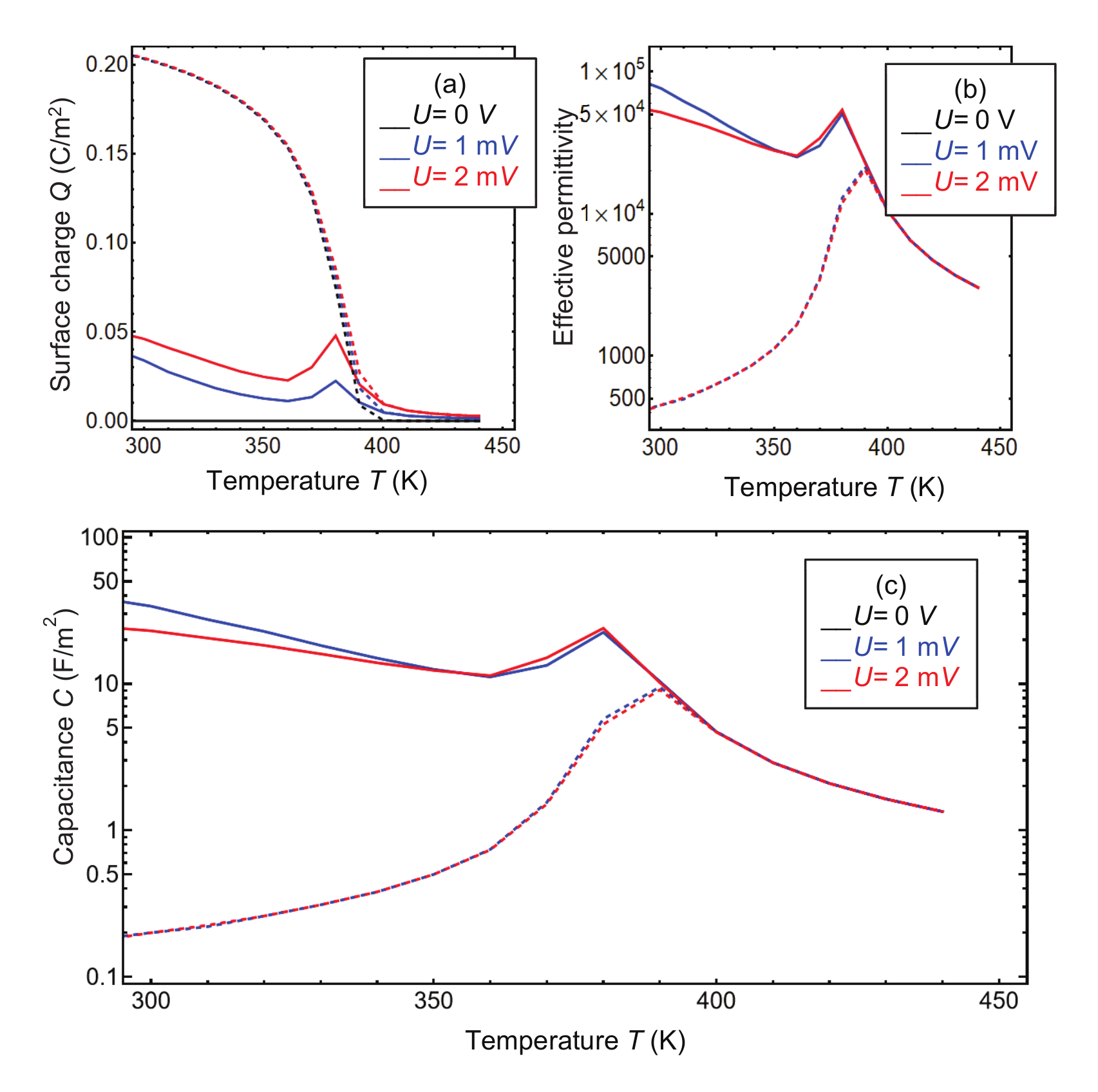}}
\caption{(Colour online) The temperature dependence of the surface charge at the electrodes (a), the effective dielectric permittivity  (b) and the capacitance (c) of the BaTiO$_3$ film placed between the electrodes calculated for different voltages $U$ indicated at the plot legends. Solid and dashed curves correspond to different initial conditions, namely poly- and single-domain states of the film, respectively. The curves are calculated for a dead layer thickness $d = 0.2$~nm and permittivity $\varepsilon _g = 81$. Film thickness is 20~nm, mismatch strain is $-0.05$\%.} \label{fig-smp2}
\end{figure}

\begin{figure}[htb]
\centerline{\includegraphics[width=0.75\textwidth]{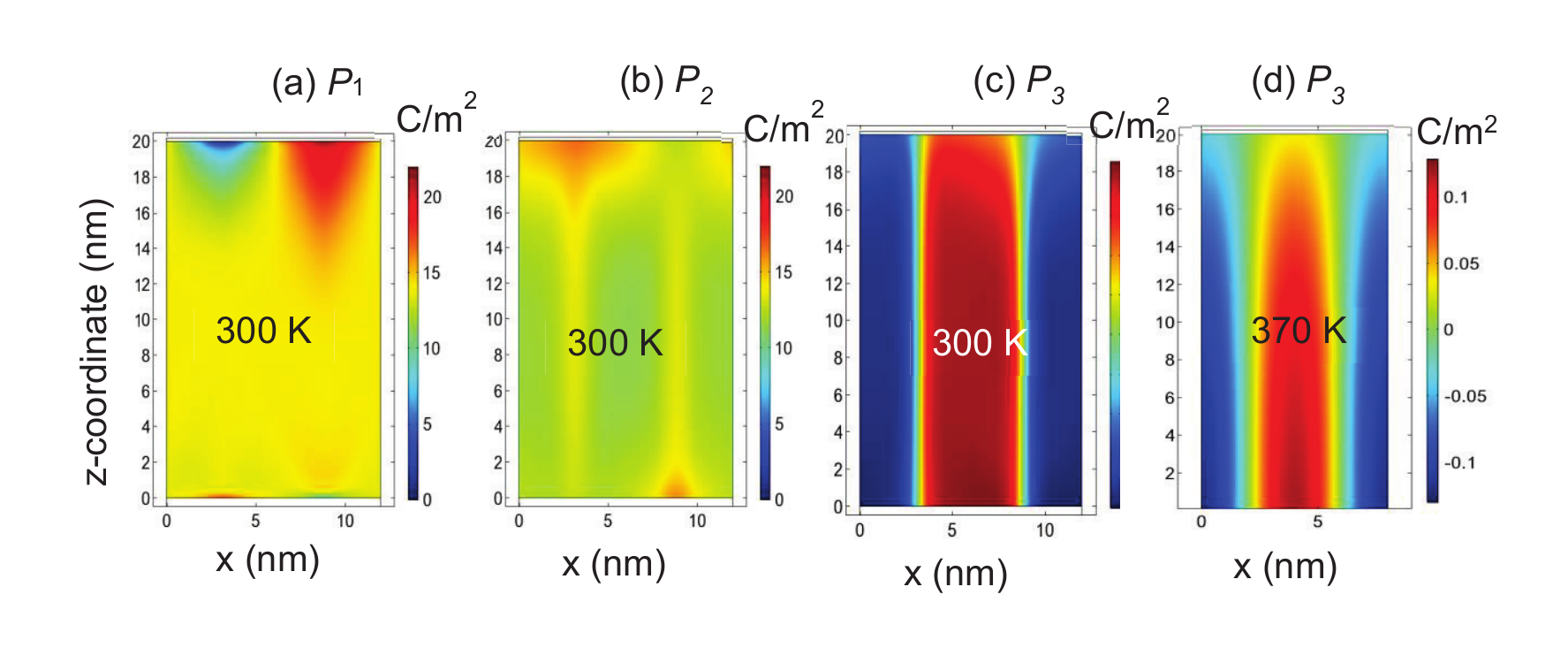}}
\caption{(Colour online) Distribution of the polarization components $P_i$ (a--c) in the cross-section of the film at 300~K, and $U = 0$. (d) One-component Ising domain wall at 370~K. Other parameters are the same as in figure~\ref{fig-smp2}.} \label{fig-smp3}
\end{figure}

The case of a single-domain BaTiO$_3$ film was used to debug the code. In this case, the thickness of the gap tends to zero, since the single-domain state may be unstable at a thickness above the critical one (less than 0.1~nm in the considered case). The dependence of the capacitance of the single-domain state on temperature is shown by dashed curves in figure~\ref{fig-smp2}, figure~\ref{fig-smp3}. BaTiO$_3$ parameters are listed in table~\ref{tbl-smp1}.

\subsection{Influence of a single domain wall on the capacitance of the ferroelectric film}

The result~(\ref{Ci}) is obvious and precise for an arbitrary distribution of $P_3 (z)$. In the case of a domain wall, the polarization depends on the transverse coordinates, $P_3 (x,z)$. Therefore, the charge density on the electrodes depends on the transverse coordinates, $\sigma (x,-d)$, primarily on the width of the wall and on the distribution of the normal polarization component on the wall. In fact, the wall-surface contact area contributes to the capacitance through these two contributions. Let us roughly estimate their influence on the capacitance of the system, based on the results of numerical modelling.

\begin{table}[!h]
	\caption{LGD coefficients and other material parameters of BaTiO$_3$.}
	\label{tbl-smp1}
	\begin{center}
		\renewcommand{\arraystretch}{0}
		\begin{tabular}{|c|c|}
			\hline
			\small Coefficient&\small Numerical value\strut\\
			\hline
			\rule{0pt}{0pt}&\\
			\hline
			\raisebox{0ex}{\small $\varepsilon _{b, e}$}
			&\small  $\varepsilon _b = 7$ (core background), $\varepsilon _e = 10$ (surrounding)\strut\\
			\hline
			\raisebox{-1.5ex}[0pt][0pt] {\small $a_i$ (in mJ/C$^2$)}
			&\small  $a_1 = 3.34(T-381)\cdot{10^5}, \alpha _T = 3.34\cdot{10^5}$ \strut\\
			&\small  $(a_1 = -2.94\cdot{10^7}$ at 298~K)\strut\\
			\hline
			\raisebox{-1.5ex}[0pt][0pt]{\small $a_{ij}$ (in m$^5$J/C$^4$)}
			&\small  $a_{11} = 4.69(T - 393) \cdot{10^6} - 2.02\cdot{10^8}, a_{12} = 3.230\cdot{10^8}$, \strut\\
			&\small   ($a_{11} = -6.71 \cdot{10^8}$ at 298~K)\strut\\
			\hline
			\raisebox{-3ex}[0pt][0pt]{\small $a_{ijk}$ (in m$^9$J/C$^6$)}
			&\small  $a_{111} = -5.52(T - 393) \cdot {10^7} + 2.76 \cdot {10^9}$,
			$a_{112} = 4.47\cdot{10^9}$, \strut\\
			&\small  $a_{123} = 4.91\cdot{10^9}$ (at 298~K $a_{111} = 82.8\cdot{10^8}$, \strut\\
			& \small  $a_{112} = 44.7\cdot{10^8}, a_{123} = 49.1\cdot{10^8})$\strut\\
			\hline
			\raisebox{-0.3ex}[0pt][0pt]{\small $Q_{ij}$ (m$^4$/C$^2$)}
			&\small  $Q_{11} = 0.11, Q_{12} = -0.043, Q_{44} = 0.059$\strut\\
			\hline
			\raisebox{-0.3ex}[0pt][0pt]{\small $s_{ij}$ (in $10^{-12}$ Pa$^{-1}$)}
			&\small  $s_{11} = 8.3, s_{12} = -2.7, s_{44} = 9.24$ \strut\\
			\hline
			\raisebox{-0.3ex}[0pt][0pt]{\small $g_{ij}$ (in $10^{-10}$ m$^3$J/C$^2$)}
			&\small  $g_{11} = 5.0, g_{12} = -0.2, g_{44} = 0.2$\strut\\
			\hline
			\raisebox{-3ex}[0pt][0pt]{\small $F_{ij}$ (in $10^{-11}$ m$^3$/C); 
				$f_{ij}$ (in V)}      &\small  $F_{11} = 2.4, F_{12} = 0.5, F_{44} = 0.06$ \strut\\ 
			&\small  (these values are recalculated from the values\strut\\
			&\small  $f_{11} = 5.1, f_{12} = 3.3, f_{44} = 0.065$ V). \strut\\
			& \small The equality $F_{44} = F_{11} - F_{12}$
			is valid in the isotropic case.\strut\\
			\hline
		\end{tabular}
		\renewcommand{\arraystretch}{1}
	\end{center}
\end{table}

As it follows from the FEM results at $U = 0$, the domain wall significantly broadens when approaching an electrically open surface $z = 0$. However, this broadening occurs in a relatively thin sub-surface layer near the top surface due to the depolarization field, and it is absent near the bottom surface due to its perfect electrical contact with the electrode~\cite{Eli09}. In the middle of the film and near the bottom electrode, which is in a perfect electrical contact with the film without any gap, a 180-degree Ising domain wall is characterized by a profile, ${P_3}\left( {x,z} \right) \approx {P_b}\tanh\left( x/L_c \right)$, where $L_c$ is the correlation length that determines the width of the wall. The same domain wall, broadened near the surface due to the depolarization field, has a different profile, ${P_3}\left( {x,z} \right) \approx P_s \left( x/L_s \right)/\left[ (x/L_s)^2 + 1 \right]$, where $L_s$ is a characteristic length, which exceeds (and sometimes significantly) the value of $L_c$. It should be noted that the value of $L_c$ significantly depends on the type of the wall and on the orientation of polarization vector.

The change in the charge density on electrodes associated with the presence of a 180-degree wall with a plane $x = 0$ can be approximated in the following way:
\begin{align}
\label{D-sigma}
\Delta \sigma \left( {x, - d} \right) \approx \frac{{{{\overline P }_3}}}{{1 + ({\varepsilon _b}/{\varepsilon _g})(d/h)}}f\left( {\frac{x}{{{L_c}}}} \right).
\end{align}
Here, we regard that $U = 0$, $f(x/L_c )$ is an odd function, equal to either $+1$ or $-1$ far away from the wall, i.e., at $\left| x \right| \gg {L_c}$. In the interval $\left| x \right| \gg {L_c}$, the function $f(x/L_c)$ changes its sign. Taking into account that the charge density contains an alternating contribution, which dominates at low voltages $U$ (when the wall is stationary), the contribution to the capacitance in the presence of a domain structure is actually made only by the difference in polarization values taken to the left and to the right of the domain wall caused by the voltage dependence of polarization, which is found from the nonlinear EL equations. These equations correspond to minimization of the LGD functional.

Under the increase of $U$, the main contribution to capacitance is made by the shift of the domain walls. Due to an increase in the effective capacitance of the domain wall region, this contribution can significantly exceed the contribution from single-domain regions. It can be estimated as follows:
\begin{align}
\label{CQU}
C = \frac{{\rd Q}}{{\rd U}} = \frac{{{\varepsilon _0}{\varepsilon _b}{\varepsilon _g}S}}{{{\varepsilon _g}h + {\varepsilon _b}d}} + \frac{{{\varepsilon _g}hS}}{{{\varepsilon _g}h + {\varepsilon _b}d}}\frac{{2{L_c}}}{L}{\chi _{DW}}\left( T \right).
\end{align}
Here, $L$ is the film length along $x$-axis, which is perpendicular to the wall plane. The function $\chi _{DW} (T)$ is the effective dielectric susceptibility of the domain wall, which, in principle, contains information about all the changes in the wall structure, including phase transitions. However, this information is indirect, and, as our FEM shows, it is not trivial to separate it from other contributions, such as transformation to a single in the external electric field (actually, departure of domain walls).

The charge at the electrodes, effective dielectric permittivity and capacitance of the BaTiO$_3$ film with thickness 20~nm are shown in figure~\ref{fig-smp2} (a), (b) and (c), respectively. A very small compressive strain $u_m = -0.05$\% is applied to the ferroelectric film, leading to the disappearance of the region of meta-stable orthorhombic phase and to the shift of the temperatures of structural and polar transitions~\cite{Per98},~\cite{Kou01}. The transition from rhombohedral (Rh) to tetragonal (Te) phase occurs at a temperature of about 360~K, and the transition from the Te to paraelectric (PE) phase occurs just above 410~K. Herein, the Te-PE transition is smeared due to the presence of internal electric field. In the case of a single-domain film, the Rh-Te phase transition appearing at 360~K is very close to the second order transition. That is why one could see a subtle kink on the dashed curves. In the case of a polydomain film, a sharp jump is seen at this temperature, that is very close to the first order phase transition. It is caused by the movement of the domain wall, since in the vicinity of a phase transition the critical field is very small and the walls move even under a tiny electric field. Thus, the ferroelectric film tends to become a single-domain in this case. The permittivity is calculated based on the capacitance of the system, obtained by differentiation of the charge density under the voltage increase. Similarly, the application of a very small electric field destroys the poly-domain structure near the point of transition to the PE phase. 

In the Rh phase, that is stable below 350~K, all three components of polarization are non-zero and change their signs when coming across the 180-degree wall (see figure~\ref{fig-smp3}). In the Te phase, that is stable above 360~K, the walls become of Ising type, i.e., only one component of the polarization exists. One of the periods of the polydomain structure is shown.

\begin{figure}[htb]
\centerline{\includegraphics[width=0.65\textwidth]{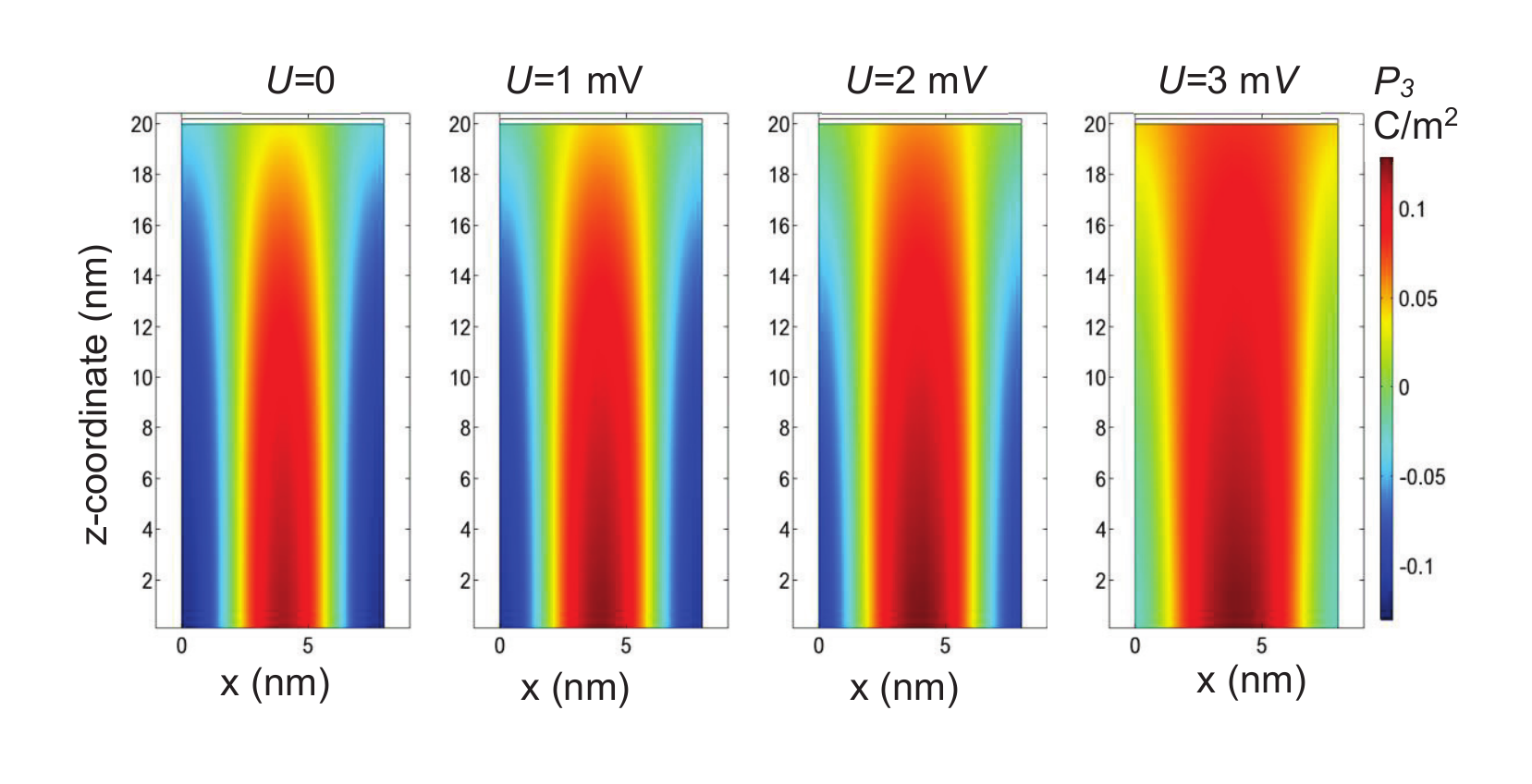}}
\caption{(Colour online) Distribution of polarization component $P_3$ in the cross-section of the film calculated for different voltage values $U = 0$, 1, 2 and 3 mV (from left to right) at 370~K. Other parameters are the same as in figure~\ref{fig-smp2}.} \label{fig-smp4}
\end{figure}

At the temperatures above 360~K, but this depends on the voltage, a giant dielectric susceptibility is associated with the movement of the domain walls under the influence of the electric field (see figure~\ref{fig-smp4}). Only one period of the polydomain structure is shown to demonstrate the trend towards the development of a single domain state (``the run-away'' or departure of the  walls). Herein, the critical field of movement of the walls depends on the temperature, and its definition is a separate task.

\section{Conclusion}

Using LGD theory and FEM, we analytically and numerically study the influence of the domain walls on the capacitance of the multiaxial ferroelectric BaTiO$_3$ film. We analyze the distinct features of the phase transitions in the domain walls, which are potentially detectable using the scanning probe capacitance microwave microscopy. 

In the Rh phase, that is stable below 350~K, all three components of polarization are non-zero and change their signs when coming across the 180-degree wall. In the Te phase, that is stable above 360~K, the walls become of Ising type, i.e., only one component of the polarization exists. Furthermore, we observe the trend towards the development of a single domain state (``the run-away'' or departure of walls). Herein, the critical field of movement of the walls depends on the temperature, and its definition is a separate task.

\newpage

%
%


\ukrainianpart

\title{Фазові переходи в сегнетоелектричних доменних стінках}
\author{Є. А. Єлісєєв\refaddr{label1}, М. Д. Глинчук\refaddr{label1}, Л. П. Юрченко\refaddr{label1}, П. Максимович\refaddr{label2}, Г.~М.~Морозовська\refaddr{label3}}
\addresses{
\addr{label1}  Інститут проблем матеріалознавства НАН України, вул. Кжижановського 3, 03142 Київ, Україна 
\addr{label2} Центр нанофазного матеріалознавства, Національна лабораторія Окріджа, Окрідж, Теннессі 37831, США
\addr{label3} Інститут фізики НАН України, просп. Науки, 03028 Київ, Україна 
}

\makeukrtitle

\begin{abstract}
\tolerance=3000%
Незважаючи на значні зусилля, існує багато невирішених фундаментальних проблем, пов’язаних із вияв\-ленням та аналізом внутрішньої структури поляризації та відповідних фазових переходів у сегнето\-електричних доменних стінках. Їх рішення може бути дуже важливим для прогресу застосування доменних стінок у наноелектроніці та пов’язаного із ним застосування у елементах пам’яті та інших інформаційних технологіях. У роботі теоретично вивчаються особливості фазових переходів у доменних стінках, які потенційно можна виявити за допомогою скануючої зондової ємнісної мікрохвильової мікроскопії. Скінченно-елементне моделювання на основі теорії Ландау-Гінзбурга-Девоншира виконано для зміни ємності, пов'язаної з рухом доменної стінки в багатоосьовому сегнетоелектрику BaTiO$_3$.
\keywords сегнетоелектрики, доменні стіни, локальні фазові переходи

\end{abstract}

\lastpage
\end{document}